\documentclass[twocolumn,showpacs,preprintnumbers,amsmath,amssymb]{revtex4}

\usepackage{graphicx}
\usepackage{dcolumn}
\usepackage{bm}
 \usepackage{graphicx} 
 \usepackage{epstopdf}
\usepackage{url}
\PassOptionsToPackage{hyphens}{url}\usepackage{hyperref}
\usepackage[usenames,dvipsnames]{color}
\usepackage[hyphenbreaks]{breakurl}

\DeclareMathOperator{\sign}{sign}

\newcommand{\pdn}{{P_{\downarrow}}}

\begin{document}


\title{Internet comments as a barometer of public opinion}

\author{Elad Oster}
\affiliation{The Racah Institute of Physics, The Hebrew University of Jerusalem, Jerusalem 91904, Israel}
\author{Erez Gilad}
\affiliation{The Unit of Nuclear Engineering, Ben-Gurion University of the Negev, Beer-Sheva 84105, Israel}
\author{Alexander Feigel}
\email{alexfeigel@gmail.com}
\affiliation{Dept. of Applied Physics, Soreq Nuclear Research Center (SNRC), Yavne 81800, Israel}


\date{\today}

\begin{abstract}

Social susceptibility is defined and analyzed using data from CNN news website. The current models of opinion dynamics, voting, and herding in closed communities are extended, and the community's response to the injection of a group with predetermined and permanent opinions is calculated. A method to estimate the values of possible response in Internet communities that follow a specific developing subject is developed. The level of social influence in a community follows from the statistics of responses ("like" and "dislike" votes) to the comments written by the members of the same community. Three real cases of developing news stories are analyzed. We suggest that Internet comments may predict the level of social response similar to a barometer that predicts the intensity of a coming storm in still calm environment.

\end{abstract}

\maketitle

In recent years, governments throughout the Arab world have been overthrown by uprisings that followed the self-immolation of a single person, Mohamed Bouazizi. Similarly, the Occupy Wall Street protest movement was triggered by a single call to action via a social network. Such cases raise an important question: How can an individual possessing no special reputation or authority mobilize an entire community by a single call to stand and fight, while large and professionally organized companies may remain unnoticed? Answering this question will help estimating the appropriate timing and the required size for an initial group to evoke a large-scale social response.

A clear and strong display of personal opinions affects the decision-making processes of others. This phenomenon of social influence may be either positive or negative. Positive social influence facilitates the correlated behavior called herding\cite{BANERJEE1992}. Herding contributes significantly to the formation of market prices\cite{Cont2000}\cite{Sznajd-Weron2002}, the results of artificial market experiments\cite{Salganik_2006}\cite{Borghesi2007}\cite{Muchnik2013}, traffic flows\cite{Helbing2001}, voting outcomes\cite{Durrett2012}\cite{Fernandez-Gracia2014}, and dynamics of social networks\cite{Sznajd2000}\cite{Galam-2003}\cite{Castellano2009}.

Acute herding phenomena, such as social revolutions or financial crises, are extremely difficult to predict, though they are evident when they occur\cite{Sornette2003}. A parameter, such as temperature in phase transitions, is required to estimate the stability of a community's opinion, i.e. the potential of a small perturbation to culminate in abrupt changes in opinion dynamics. 
Therefore, to understand the population dynamics prior to a possible transition, it is important to develop a quantitative analysis of herding as a function of time.

Internet communities are of special interest for the analysis of the herding phenomenon. Individual opinions are widely exposed in binary form of "like" and "dislike" votes ("likes" and "dislikes") over Internet news websites and via social networks. 
The data span any important event and expose millions of opinions\cite{Bond2012}. Simultaneous analysis of a developing news story and the corresponding herding in relevant Internet communities may provide a unique opportunity to study the opinion dynamics in a population as it approaches a critical point and becomes unstable. To the best of our knowledge, the definition and evaluation of the temporal dynamics of herding phenomenon in Internet communities remains a challenge.

In this Article, we estimate the social influence as a function of time in Internet communities that followed any of the following three news stories reported on the CNN website: the Zimmerman trial, Iran Nuclear Negotiations, and the US government shutdown of 2013. 
We show continuous herding dynamics in all three cases and 
significant amplification of social influence near the verdict announcement in the Zimmerman case. 
The method we propose allows for the quantitative estimation of a community response to the injection of a group of non-responsive individuals with predefined opinion. This quantitative analysis is possible due to our novel approach to herding as the conditional probabilities to agree or disagree with other people's opinions. This approach differs from the generally accepted treatment of herding as a topology of social interactions' network\cite{Cont2000}.

\begin{figure}
	\resizebox{0.4\textwidth}{!}{\includegraphics{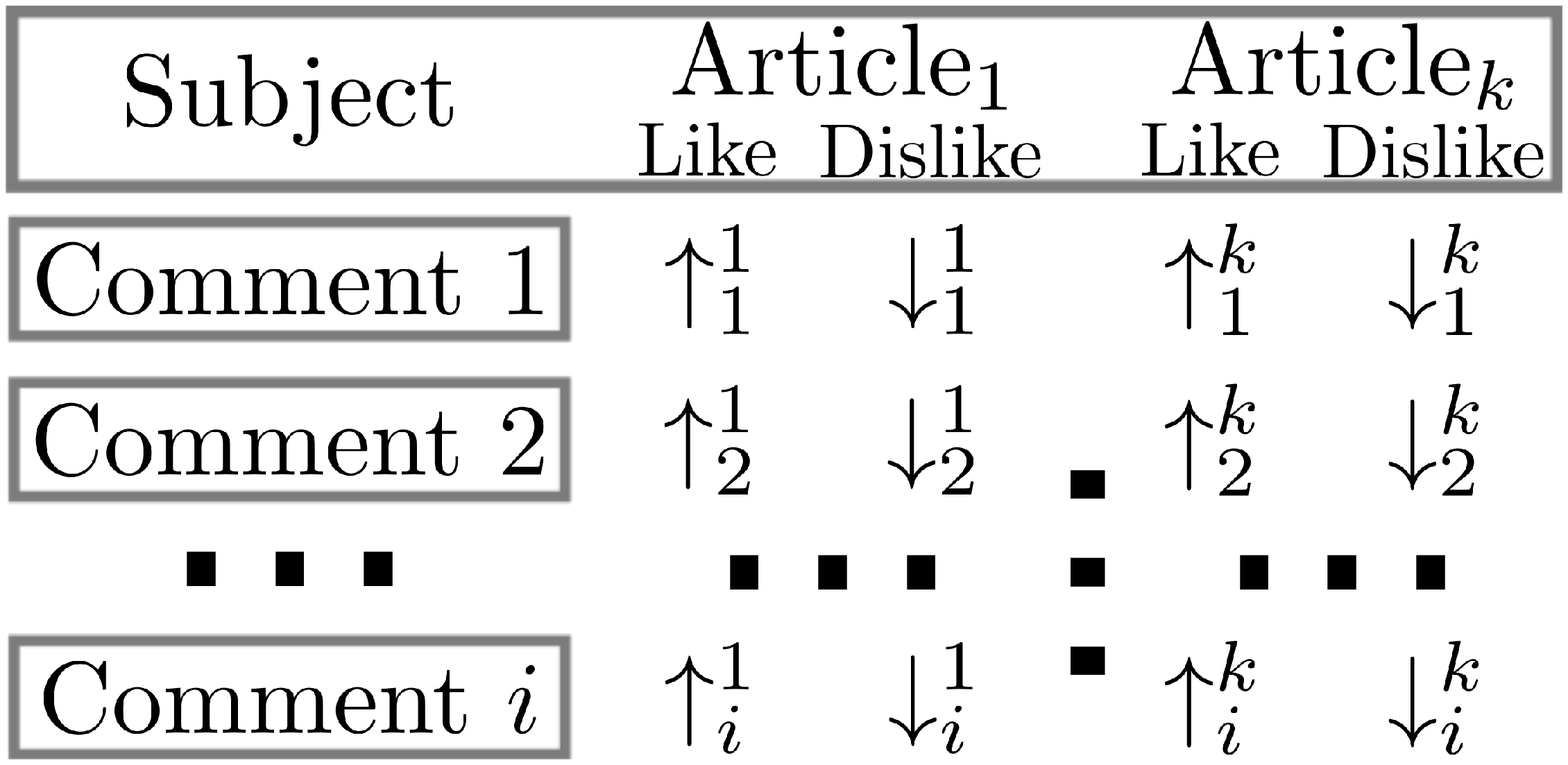}}
	\caption{Internet news and social influence. Consider articles that follow some developing news story. Contrary to printed newspapers, Internet news websites open some articles for commentary by the public and for expressing like or dislike votes for each comment. Comments, together with their likes and dislikes, are written and voted for from both supporters and opponents of the article’s statements. Quantitative data of likes $(\uparrow )$ and dislikes $(\downarrow )$ reveal the conditional probabilities of individual community members to respond positively or negatively to others’ opinions. These conditional probabilities reflect the level of social influence in the community and allow monitoring the temporal dependence of the level of social influence by following articles on the same subject from different dates.
}
	\label{fig:CNN_analysis_diagram}
\end{figure}

To estimate social susceptibility, we use a specific type of Internet news discussion. Some Internet news websites provide a commentary section where readers can comment and vote (i.e., like or dislike) other readers' comments (see Fig.~\ref{fig:CNN_analysis_diagram}). 
A reader can usually vote for any number of comments, with the restriction of one vote per comment. These data constitute a natural large scale social experiment where the population responds to some external signal (i.e., a comment). A comment, however, is not completely external, but rather created by a community member who responds to the comments of other community members. Consequently, statistics of Internet comments and responses can be used as a measure for mean field opinion dynamics of the corresponding community.

Consider a large population of $N$ individuals who are debating on a subject $S$ and continuously voting in favor of $S$ (up $\uparrow$) or against it (down $\downarrow$). The debate process implies that individuals may change their vote in time. In our model, the interaction between individual $i$ and any other randomly selected individual $j$ is expressed by the fact that the probability per contact of individual $i$ to vote down ($\pdn^{ij}$) depends on the vote of individual $j$. This conditional probability is given by \cite{Feigel2008}:
\begin {eqnarray}
\pdn^{ij} =
	\begin{cases} 
		\alpha_{ij} & \mbox{if } s_j =1 \\
		\beta_{ij}  & \mbox{if } s_j = 0 
		\end{cases} 
		\quad
	= \alpha_{ij} s_j + \beta_{ij} (1-s_j),
\label{eq:conditional probability 1}
\end {eqnarray}
where $s_j$ is the vote of individual $j$ ($s_j = 1$ for up vote and $s_j = 0$ for down vote) and parameter $\alpha_{ij}$ ($\beta_{ij}$) is the probability per contact of individual $i$ voting down given individual $j$ is voting up (down), respectively, regardless of the vote of individual $i$ prior to the interaction with individual $j$.

In a well mixed homogeneous population, where the number of contacts per individual is $k_i = N$ and $(\alpha_{ij}, \beta_{ij})=(\alpha, \beta)$, the probability of an individual to vote down is
\begin {equation}\label{eq:conditional probability 3}
\pdn\equiv\pdn^i =  \frac{1}{k_i} \sum_{j=1}^{k_i}{\pdn^{ij}} = \frac{1}{N} \sum_{j=1}^N {\left[\alpha s_j + \beta (1-s_j)\right]}.
\end {equation}
Defining $\gamma\equiv \frac{1}{N}\sum _{i=1}^N {s_i}$ as the mean fraction of individuals who vote up, and noting that mean field assumptions imply $\gamma = P_{\uparrow} = 1-P_{\downarrow}$, Eq.~\eqref{eq:conditional probability 3} may be written as  
\begin {equation}\label{eq:Pdown}
\pdn=1-\gamma=\gamma\alpha + (1-\gamma)\beta ,
\end {equation}
resulting in a steady state expression for $\gamma$ (the "public opinion") as a function of conditional probabilities
\begin {eqnarray}\label{eq:mean-field-gamma}
\gamma=\frac{1-\beta}{1+\alpha-\beta}.
\end {eqnarray}

\begin{figure}
\resizebox{0.5\textwidth}{!}{\includegraphics{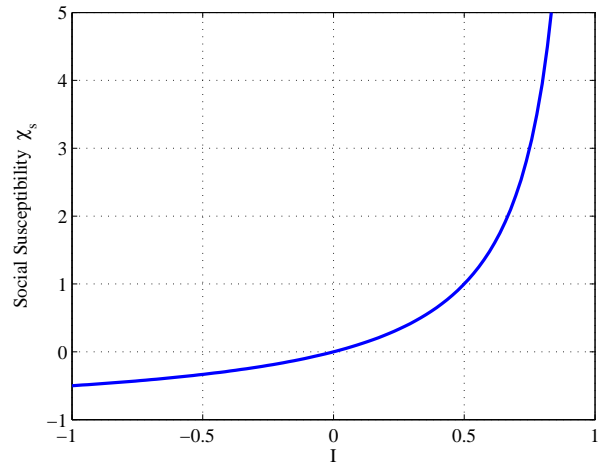}}
\caption{The social susceptibility $\chi_s$ as a function of herding parameter $I$. Social susceptibility is a measure of how many individuals follow a single one who changes his or her opinion. Thus $\chi_s>>1$ ($I\approx 1$) makes possible significant social transitions that are initiated by a small group. 
Social influence vanishes if $\chi_s=0$. The case of $\chi_s<0$ corresponds to the populations with negative (antagonistic) social influence.}
\label{fig:SAF}
\end{figure}
In order to measure social influence, consider a population of $N$ individuals characterized by $(\alpha, \beta)$, which is perturbed by applying the specific value of mean vote $\gamma_\rho$ to a fraction $\rho\in[0,1]$ of the population. The new mean vote of the population $\gamma_{n}$ is given by
\begin{eqnarray}
	\label{eq:r1}
	\gamma_{n}=(1-\rho)[\gamma_{n}(1-\alpha)+(1-\gamma_{n})(1-\beta)]+\rho\gamma_{\rho}.
\end{eqnarray}
The response function of the population $R(\rho)$ is defined by the fraction of players who flip votes \emph{in response} to the perturbation, i.e. outside the perturbation group.
An explicit expression for $R(\rho)$ is obtained using Eq.~\eqref{eq:r1}
\begin{equation}\label{eq:r2}
R(\rho)=\frac{\sign (\gamma_{\rho}-\gamma)(\gamma_{n}-\gamma)-\rho|\gamma_{\rho}-\gamma|}{1-\rho}=\frac{I\rho}{1-I(1-\rho)},
\end{equation}
where $I=\beta-\alpha$. Obviously, the population response function $R(\rho)$ is zero for $\rho=0$ and for $\alpha=\beta$. 

The herding parameter $I\in[-1,1]$ is a measure of the social influence of one individual on others, because $I=\beta-\alpha$ is the difference of conditional probabilities for correlated and anti-correlated behaviors, see \ref{eq:conditional probability 1}. It is similar to herding or percolation parameter $0\leq c\leq 1$ from~\cite{Cont2000}. However, since our definition of the herding parameter $I$ accounts for both positive and negative social influence, it is better suited for analyzing opinion dynamics in binary vote communities.

The social susceptibility $\chi_s$, is defined as
\begin{equation}\label{eq:chi1}
\chi_s \equiv \left.\frac{\partial R}{\partial\rho}\right|_{\rho=0}=\frac{I}{1-I},
\end{equation}
and is the average size of a group whose members follow the change of opinion of a single member (not including the initiating member itself). The size of the perturbation group $\rho_{crit}$ required to convert a population $(\alpha,\beta)$ to the mean vote of the perturbed group $\gamma_{\rho}$ (including the polarized cases $\gamma_{\rho}=1,0$) is obtained by substituting $\gamma_n = \gamma_{\rho}$ in Eq.~\eqref{eq:r2} and using Eq.~\eqref{eq:chi1} 
\begin{equation}
\label{eq:chi2}
\rho_{crit}=\frac{1}{\chi_s}\frac{|\Delta\gamma|}{(1-|\Delta\gamma|)},
\end{equation} 
where $\Delta\gamma=\gamma_\rho-\gamma$. 

Conditional probabilities $(\alpha,\beta)$ define herding $I$, which in turn defines the social stability of the community. To calculate conditional probabilities $\alpha$ and $\beta$ as a function of likes $\uparrow_i$ and dislikes $\downarrow_i$ votes for comment $i$ of article $k$ (see Fig.~\ref{fig:CNN_analysis_diagram}), we assume that voters’ and commentators’ populations are equivalent and that the number of comments and votes is large enough to apply mean field assumption. Consequently, the probabilities for a commentator and a voter to be in favor of the article subject $S$ are both equal to $\gamma$. Therefore, the comments should consist of two groups with opposite opinions and relative sizes $\gamma$ and $1-\gamma$, respectively. 

According to the definition of the conditional probabilities (Eq.~\ref{eq:conditional probability 1}), the ratio between likes and all responses (likes and dislikes) for a positive comment (to $S$) is $1-\alpha$. However, this ratio for a negative comment (to $S$) equals $\beta$ since expressing a like vote for a negative comment is equivalent to expressing a dislike vote for the article subject $S$ commented upon. Consequently, the probability of a dislike vote for a comment $P_{\mathrm{dislike}}$ is different from the probability $P_\downarrow$ to dislike subject $S$, as defined in (Eq.~\ref{eq:Pdown}). Therefore, the probability of a dislike vote for a comment is:
\begin{eqnarray}
  \label{eq:comdown}
  P_{\mathrm{dislike}} = \alpha \gamma + (1-\beta)(1-\gamma) = 2(1-\gamma)\gamma.
\end{eqnarray}
The result is invariant under the transformation $\gamma\rightarrow 1-\gamma$, reflecting the uncertainty regarding the opinion of the Internet article itself. Hence, the division of the comments into two groups with contrasting opinions does not reveal the opinions themselves. 
Since $\chi_s$ is invariant under the transformation $\gamma\rightarrow 1-\gamma$, we arbitrarily chose $\gamma>0.5$. An interesting consequence of Eq.~\eqref{eq:comdown} is that $P_{\mathrm{dislike}}<0.5$, i.e. comments cannot include only dislikes because the community cannot dislike its own opinion.

Calculating of $\alpha$, $\beta$ and $\gamma$ of the community proceeds through iterations. First, all comments are sorted by their like vote fraction. Then, at each step $n$, the comments are divided into two groups with ratio of $\gamma^n$ and $1-\gamma^n$ according to their like vote fraction, where group $L$ receives the $\gamma^n$ comments with the highest like vote fraction and group $D$ receives all other comments.
The population characteristic parameters $\alpha^n$ and $\beta^n$ are then calculated according to: 
\begin{eqnarray} 
	 1- \alpha^n &=& \sum\limits_{i \in L}{\frac{\uparrow_i}{\uparrow_i+\downarrow_i}},\nonumber \\
	 \beta^n &=& \sum\limits_{i \in D}{\frac{\uparrow_i}{\uparrow_i+\downarrow_i}}.
\end{eqnarray}
A new population mean vote $\gamma^{n+1}$ is calculated using the values of $\alpha^n$ and $\beta^n$: 
\begin {eqnarray}
	 \gamma^{n+1}=\frac{1-\beta^n}{1+\alpha^n-\beta^n}.
\end {eqnarray}
The process is repeated until the convergence of $\alpha^n$, $\beta^n$ and $\gamma^n$. 	 

\begin{table}[ht] 
	\caption{The results of social susceptibility calculation for 12 CNN articles from different dates that cover three different events. For each article, the values of conditional probabilities $(\alpha,\beta)$ and social susceptibility $\chi_s$ were calculated.} 
	\centering 
	\begin{tabular}{c l c c c c c} 
		\hline\hline 
		& Article's topic & Publish Date & ~~$\alpha$~~ & ~~$\beta$~~ & ~~$\gamma$~~ & $\chi_s$ \\ [0.5ex] 
		\hline 
		1 & Zimmerman Trial & 24/06/13 & 0.14 & 0.53 & 0.77 & 0.63 \\ 
		2 & Zimmerman Trial & 05/07/13 & 0.12 & 0.38 & 0.84 & 0.34\\ 
		3 & Zimmerman Trial & 12/07/13 & 0.08 & 0.47 & 0.87 & 0.64\\ 
		4 & Zimmerman Trial & 13/07/13 & 0.04 & 0.57 & 0.91 & 1.10\\ 
		5 & Zimmerman Trial & 17/07/13 & 0.04 & 0.67 & 0.89 & 1.70\\ 
		6 & Zimmerman Trial & 25/07/13 & 0.06 & 0.77 & 0.79 & 2.39\\ 
		\hline
		7 & Iran Nuclear Program & 25/10/13 & 0.16 & 0.58 & 0.72 & 0.74\\ 
		8 & Iran Nuclear Program  & 23/11/13 & 0.15 & 0.56 & 0.75 & 0.70\\ 
		9 & Iran Nuclear Program  & 24/11/13 & 0.15 & 0.59 & 0.73 & 0.78\\
		\hline
		10 & US Govt. Shutdown & 01/10/13 & 0.16 & 0.51 & 0.75 & 0.53\\
		11 & US Govt. Shutdown & 02/10/13 & 0.13 & 0.51 & 0.79 & 0.63\\
		12 & US Govt. Shutdown & 02/10/13 & 0.09 & 0.48 & 0.85 & 0.62\\ [1ex] 
		\hline 
		\hline
	\end{tabular} 
	\label{table:cnn} 
\end{table} 

\begin{figure}
	\resizebox{0.5\textwidth}{!}{\includegraphics{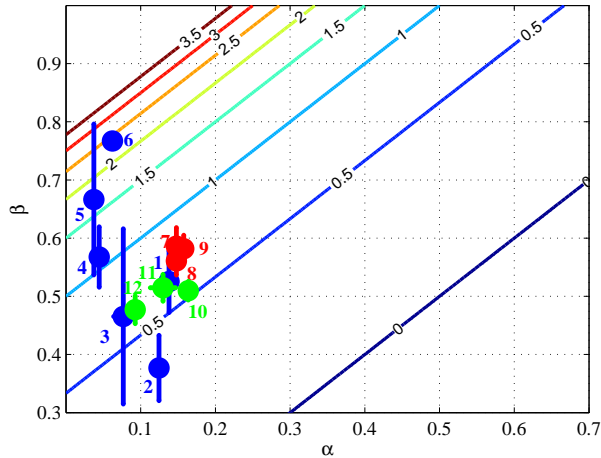}}
	\caption{Social susceptibility $\chi_s$ as a function of conditional probabilities $(\alpha,\beta)$ together with the states of different Internet communities according to the analysis of CNN website articles (numbered dots). The numbers correspond to the cases of Table \ref{table:cnn} that cover the Zimmerman trail, Iran Nuclear Program agreement, and US Government shutdown.}
	\label{fig:CNN_analysis}
\end{figure}

The formalism of the analysis of the social influence presented above is applied to news articles published on the CNN website that discuss three different topics. 
The first story includes six articles, published between June 24th and July 25th, 2013, covering the George Zimmerman Trial~\cite{Zimmerman-1,Zimmerman-2,Zimmerman-3,Zimmerman-4,Zimmerman-5,Zimmerman-6}. These articles cover the legal proceeding, the verdict, and the post-verdict jurors’ opinions about the trial. The second story includes three articles, published between October 25th and November 25th, 2013, covering the negotiations and signing of the Geneva interim agreement on the Iranian nuclear program ~\cite{Iran-1,Iran-2,Iran-3}. The third story includes three articles, published on October 1st and 2nd, 2013, covering the US federal government shutdown of that year~\cite{Shudown-1,Shudown-2,Shudown-3}. These articles cover the first day of the shutdown and the White House failing efforts to end it. The results of these analyses are presented in Table~\ref{table:cnn} and in Fig.~\ref{fig:CNN_analysis}.

\begin{figure}
	\resizebox{0.5\textwidth}{!}{\includegraphics{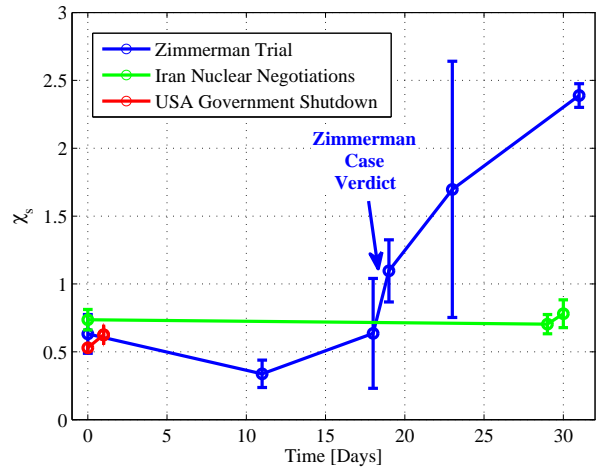}}
	\caption{The social susceptibility $\chi_s$ as a function of time.
	Social susceptibility remains almost constant in the cases of US Government shutdown and Iran Nuclear Program negotiations. This preservation of community state is surprising because there is no reason why the opinion of a population should remain the same for short or for long periods. Even more interesting, however, is that social susceptibility in the case of the Zimmerman trial changes after the verdict is announced on July 13, 2013. Significant social transition becomes possible 
	after the announcement of the verdict.}
	\label{fig:CNN_SAF_and_RHO}
\end{figure}

All three cases exhibit a continuous dynamics in $(\alpha,\beta)$ space, as shown in Fig.~\ref{fig:CNN_analysis}. This result is interesting considering that the analysis is applied to different articles, covering different stories, spanning from days to months. It indicates a slow change of opinions in the community.

In the Iran Nuclear Program and US Government shutdown cases, the population's characteristic parameters $(\alpha,\beta)$ are constant, although they correspond to different CNN articles and, in the case of the Iran Nuclear Program, span one month. This result may also indicate the absence of special events during the observation period.

The social susceptibility level in the Zimmerman trial case changes near the verdict announcement. In the period prior to the verdict day (points 1-3 in Fig.~\ref{fig:CNN_analysis}),  the level of social susceptibility in the population remains almost constant and similar to the social susceptibility in the other cases (i.e., $\chi_s\sim 0.5$), despite the changes in $\alpha$ and $\beta$. From the verdict day on (points 4-6), the social susceptibility in the community grows rapidly and the population approaches the singular point $(\alpha,\beta)\rightarrow (0,1)$. It is out of the scope of this work to interpret social phenomena, though the results demonstrate that our method allows to observe the otherwise hidden herding level in a community together with its response to social triggers.

The limitations of our work include the absence of external force, i.e. government control, and lack of interaction topology constrains, such as the prevalence of near-neighbors interactions. Omitting topological constraints seems to be justified in Internet communities. The same is true regarding forces that shape opinion or add weight to some opinion, such as government control or mass media. We assume that the Internet is still a free zone. The model can be extended to include such force, though there is no clear way to quantify it. 

Shortly after the data collection phase for this work was completed, the CNN website has changed its comments policy and the dislike count per comment is no longer displayed. This change made the CNN website articles and comments unsuitable for the above comment analysis procedure, since the main assumption underlying our model, i.e., that both like and dislike vote counts are available to all individuals in the population, is no longer valid. This study demonstrates the potential of \emph{both} like/dislike votes in estimating the social state of a community and may contribute to the evolving formation of the Internet news format.

To conclude, the developed tools for social influence in Internet communities reveal the previously hidden level of herding and social influence as a function of time in  populations. In addition, this work provides a measure for the stability of public opinion in a community and for the size of a group capable to cause critical change in average opinion. The presented method can be compared with other methods and can be extended to other fields such as financial markets\cite{Oster2015}. Therefore, this work enables an intriguing comparison of the herding in the same community calculated from different sources, such as Internet news and financial markets.


\begin{thebibliography}{28}
\expandafter\ifx\csname natexlab\endcsname\relax\def\natexlab#1{#1}\fi
\expandafter\ifx\csname bibnamefont\endcsname\relax
  \def\bibnamefont#1{#1}\fi
\expandafter\ifx\csname bibfnamefont\endcsname\relax
  \def\bibfnamefont#1{#1}\fi
\expandafter\ifx\csname citenamefont\endcsname\relax
  \def\citenamefont#1{#1}\fi
\expandafter\ifx\csname url\endcsname\relax
  \def\url#1{\texttt{#1}}\fi
\expandafter\ifx\csname urlprefix\endcsname\relax\def\urlprefix{URL }\fi
\providecommand{\bibinfo}[2]{#2}
\providecommand{\eprint}[2][]{\url{#2}}

\bibitem[{\citenamefont{Banerjee}(1992)}]{BANERJEE1992}
\bibinfo{author}{\bibfnamefont{A.}~\bibnamefont{Banerjee}},
  \bibinfo{journal}{Quant. J. of Econ.} \textbf{\bibinfo{volume}{107}},
  \bibinfo{pages}{797} (\bibinfo{year}{1992}).

\bibitem[{\citenamefont{Cont and Bouchaud}(2000)}]{Cont2000}
\bibinfo{author}{\bibfnamefont{R.}~\bibnamefont{Cont}} \bibnamefont{and}
  \bibinfo{author}{\bibfnamefont{J.}~\bibnamefont{Bouchaud}},
  \bibinfo{journal}{MacroEcon. Dyn.} \textbf{\bibinfo{volume}{4}},
  \bibinfo{pages}{170} (\bibinfo{year}{2000}).

\bibitem[{\citenamefont{Sznajd-Weron and Weron}(2002)}]{Sznajd-Weron2002}
\bibinfo{author}{\bibfnamefont{K.}~\bibnamefont{Sznajd-Weron}}
  \bibnamefont{and} \bibinfo{author}{\bibfnamefont{R.}~\bibnamefont{Weron}},
  \bibinfo{journal}{Int. J. of Mod. Phys. C} \textbf{\bibinfo{volume}{13}},
  \bibinfo{pages}{115} (\bibinfo{year}{2002}).

\bibitem[{\citenamefont{Salganik et~al.}(2006)\citenamefont{Salganik, Dodds,
  and Watts}}]{Salganik_2006}
\bibinfo{author}{\bibfnamefont{M.~J.} \bibnamefont{Salganik}},
  \bibinfo{author}{\bibfnamefont{P.~S.} \bibnamefont{Dodds}}, \bibnamefont{and}
  \bibinfo{author}{\bibfnamefont{D.~J.} \bibnamefont{Watts}},
  \bibinfo{journal}{Science} \textbf{\bibinfo{volume}{311}},
  \bibinfo{pages}{854} (\bibinfo{year}{2006}).

\bibitem[{\citenamefont{Borghesi and Bouchaud}(2007)}]{Borghesi2007}
\bibinfo{author}{\bibfnamefont{C.}~\bibnamefont{Borghesi}} \bibnamefont{and}
  \bibinfo{author}{\bibfnamefont{J.-P.} \bibnamefont{Bouchaud}},
  \bibinfo{journal}{Quality \& Quantity} \textbf{\bibinfo{volume}{41}},
  \bibinfo{pages}{557} (\bibinfo{year}{2007}).

\bibitem[{\citenamefont{Muchnik et~al.}(2013)\citenamefont{Muchnik, Aral, and
  Taylor}}]{Muchnik2013}
\bibinfo{author}{\bibfnamefont{L.}~\bibnamefont{Muchnik}},
  \bibinfo{author}{\bibfnamefont{S.}~\bibnamefont{Aral}}, \bibnamefont{and}
  \bibinfo{author}{\bibfnamefont{S.~J.} \bibnamefont{Taylor}},
  \bibinfo{journal}{Science} \textbf{\bibinfo{volume}{341}},
  \bibinfo{pages}{647} (\bibinfo{year}{2013}).

\bibitem[{\citenamefont{Helbing}(2001)}]{Helbing2001}
\bibinfo{author}{\bibfnamefont{D.}~\bibnamefont{Helbing}},
  \bibinfo{journal}{Rev. of Mod. Phys.} \textbf{\bibinfo{volume}{73}},
  \bibinfo{pages}{1067} (\bibinfo{year}{2001}).

\bibitem[{\citenamefont{Durrett et~al.}(2012)\citenamefont{Durrett, Gleeson,
  Lloyd, Mucha, Shi, Sivakoff, Socolar, and Varghese}}]{Durrett2012}
\bibinfo{author}{\bibfnamefont{R.}~\bibnamefont{Durrett}},
  \bibinfo{author}{\bibfnamefont{J.~P.} \bibnamefont{Gleeson}},
  \bibinfo{author}{\bibfnamefont{A.~L.} \bibnamefont{Lloyd}},
  \bibinfo{author}{\bibfnamefont{P.~J.} \bibnamefont{Mucha}},
  \bibinfo{author}{\bibfnamefont{F.}~\bibnamefont{Shi}},
  \bibinfo{author}{\bibfnamefont{D.}~\bibnamefont{Sivakoff}},
  \bibinfo{author}{\bibfnamefont{J.~E.~S.} \bibnamefont{Socolar}},
  \bibnamefont{and} \bibinfo{author}{\bibfnamefont{C.}~\bibnamefont{Varghese}},
  \bibinfo{journal}{Proc. of Nat. Acad. of Science}
  \textbf{\bibinfo{volume}{109}}, \bibinfo{pages}{3682} (\bibinfo{year}{2012}).

\bibitem[{\citenamefont{Fernandez-Gracia
  et~al.}(2014)\citenamefont{Fernandez-Gracia, Suchecki, Ramasco, San~Miguel,
  and Eguiluz}}]{Fernandez-Gracia2014}
\bibinfo{author}{\bibfnamefont{J.}~\bibnamefont{Fernandez-Gracia}},
  \bibinfo{author}{\bibfnamefont{K.}~\bibnamefont{Suchecki}},
  \bibinfo{author}{\bibfnamefont{J.~J.} \bibnamefont{Ramasco}},
  \bibinfo{author}{\bibfnamefont{M.}~\bibnamefont{San~Miguel}},
  \bibnamefont{and} \bibinfo{author}{\bibfnamefont{V.~M.}
  \bibnamefont{Eguiluz}}, \bibinfo{journal}{Phys. Rev. Lett.}
  \textbf{\bibinfo{volume}{112}} (\bibinfo{year}{2014}).

\bibitem[{\citenamefont{Sznajd-Weron and Sznajd}(2000)}]{Sznajd2000}
\bibinfo{author}{\bibfnamefont{K.}~\bibnamefont{Sznajd-Weron}}
  \bibnamefont{and} \bibinfo{author}{\bibfnamefont{J.}~\bibnamefont{Sznajd}},
  \bibinfo{journal}{Int. J. of Mod. Phys. C} \textbf{\bibinfo{volume}{11}},
  \bibinfo{pages}{1157} (\bibinfo{year}{2000}).

\bibitem[{\citenamefont{Galam}(2003)}]{Galam-2003}
\bibinfo{author}{\bibfnamefont{S.}~\bibnamefont{Galam}},
  \bibinfo{journal}{Phys. A: Stat. Mech.} \textbf{\bibinfo{volume}{320}},
  \bibinfo{pages}{571 } (\bibinfo{year}{2003}).

\bibitem[{\citenamefont{Castellano et~al.}(2009)\citenamefont{Castellano,
  Fortunato, and Loreto}}]{Castellano2009}
\bibinfo{author}{\bibfnamefont{C.}~\bibnamefont{Castellano}},
  \bibinfo{author}{\bibfnamefont{S.}~\bibnamefont{Fortunato}},
  \bibnamefont{and} \bibinfo{author}{\bibfnamefont{V.}~\bibnamefont{Loreto}},
  \bibinfo{journal}{Rev. of Mod. Phys.} \textbf{\bibinfo{volume}{81}},
  \bibinfo{pages}{591} (\bibinfo{year}{2009}).

\bibitem[{\citenamefont{Sornette}(2003)}]{Sornette2003}
\bibinfo{author}{\bibfnamefont{D.}~\bibnamefont{Sornette}},
  \bibinfo{journal}{Phys. Reports} \textbf{\bibinfo{volume}{378}},
  \bibinfo{pages}{1} (\bibinfo{year}{2003}).

\bibitem[{\citenamefont{Bond et~al.}(2012)\citenamefont{Bond, Fariss, Jones,
  Kramer, Marlow, Settle, and Fowler}}]{Bond2012}
\bibinfo{author}{\bibfnamefont{R.~M.} \bibnamefont{Bond}},
  \bibinfo{author}{\bibfnamefont{C.~J.} \bibnamefont{Fariss}},
  \bibinfo{author}{\bibfnamefont{J.~J.} \bibnamefont{Jones}},
  \bibinfo{author}{\bibfnamefont{A.~D.} \bibnamefont{Kramer}},
  \bibinfo{author}{\bibfnamefont{C.}~\bibnamefont{Marlow}},
  \bibinfo{author}{\bibfnamefont{J.~E.} \bibnamefont{Settle}},
  \bibnamefont{and} \bibinfo{author}{\bibfnamefont{J.~H.}
  \bibnamefont{Fowler}}, \bibinfo{journal}{Nature}
  \textbf{\bibinfo{volume}{489}}, \bibinfo{pages}{295} (\bibinfo{year}{2012}).

\bibitem[{\citenamefont{Feigel}(2008)}]{Feigel2008}
\bibinfo{author}{\bibfnamefont{A.}~\bibnamefont{Feigel}}, \bibinfo{journal}{J.
  of Theor. Biol.} \textbf{\bibinfo{volume}{254}}, \bibinfo{pages}{768}
  (\bibinfo{year}{2008}).

\bibitem[{Zim({\natexlab{a}})}]{Zimmerman-1}
\emph{\bibinfo{title}{Zimmerman opening statements: Expletives and a
  knock-knock joke}},
  \urlprefix\url{http://edition.cnn.com/2013/06/24/justice/zimmerman-trial/}.

 \bibitem[{Zim({\natexlab{b}})}]{Zimmerman-2}
 \emph{\bibinfo{title}{In zimmerman trial, it's a jury of millions}},
   \urlprefix\url{http://edition.cnn.com/2013/07/05/justice/zimmerman-jury-millions/}.

\bibitem[{Zim({\natexlab{c}})}]{Zimmerman-3}
\emph{\bibinfo{title}{After days of court drama, jurors set to resume
  deliberating zimmerman's fate}},
  \urlprefix\url{http://edition.cnn.com/2013/07/12/justice/zimmerman-trial/}.

\bibitem[{Zim({\natexlab{d}})}]{Zimmerman-4}
\emph{\bibinfo{title}{George zimmerman found not guilty of murder in trayvon
  martin's death}},
  \urlprefix\url{http://edition.cnn.com/2013/07/13/justice/zimmerman-trial/}.

\bibitem[{Zim({\natexlab{e}})}]{Zimmerman-5}
\emph{\bibinfo{title}{Exclusive: Juror pushes for new laws following zimmerman
  trial}},
  \urlprefix\url{http://edition.cnn.com/2013/07/17/justice/zimmerman-verdict-aftermath/}.

\bibitem[{Zim({\natexlab{f}})}]{Zimmerman-6}
\emph{\bibinfo{title}{Zimmerman juror to abc: He 'got away with murder'}},
  \urlprefix\url{http://edition.cnn.com/2013/07/25/justice/zimmerman-juror-b29-interview/}.

\bibitem[{Ira({\natexlab{a}})}]{Iran-1}
\emph{\bibinfo{title}{Nuclear group: Time iran would need to make uranium for a
  bomb 'too short'}},
  \urlprefix\url{http://edition.cnn.com/2013/10/25/world/meast/iran-nuclear-report/}.

\bibitem[{Ira({\natexlab{b}})}]{Iran-2}
\emph{\bibinfo{title}{Obama: Iran nuclear deal limits ability to create nuclear
  weapons}},
  \urlprefix\url{http://edition.cnn.com/2013/11/23/world/meast/iran-nuclear-talks-geneva/}.

\bibitem[{Ira({\natexlab{c}})}]{Iran-3}
\emph{\bibinfo{title}{3-decade gridlock broken: The nuclear deal with iran in
  geneva}},
  \urlprefix\url{http://edition.cnn.com/2013/11/24/world/meast/iran-nuclear-deal/}.

\bibitem[{Shu({\natexlab{a}})}]{Shudown-1}
\emph{\bibinfo{title}{Latest house bid fails as bitter back-and-forth over
  government shutdown rages}},
  \urlprefix\url{http://edition.cnn.com/2013/10/01/politics/government-shutdown/}.

\bibitem[{Shu({\natexlab{b}})}]{Shudown-2}
\emph{\bibinfo{title}{Progress? obama invites congressional leaders for talks
  on shutdown}}.

\bibitem[{Shu({\natexlab{c}})}]{Shudown-3}
\emph{\bibinfo{title}{No end in sight to government shutdown after
  'unproductive' white house meeting}},
  \urlprefix\url{http://edition.cnn.com/2013/10/02/politics/government-shutdown/}.

\bibitem[{\citenamefont{Oster and Feigel}(2015)}]{Oster2015}
\bibinfo{author}{\bibfnamefont{E.}~\bibnamefont{Oster}} \bibnamefont{and}
  \bibinfo{author}{\bibfnamefont{A.}~\bibnamefont{Feigel}},
  \bibinfo{journal}{submitted}  (\bibinfo{year}{2015}).

\end{thebibliography}

\newpage

\section{Supplementary Material}

Here we include the detailed procedure to obtain the social influence parameter from Internet discussion data. The algorithm’s input are two vectors containing the number of likes and dislikes each comments received, $\uparrow_i$ and $\downarrow_i$, respectively (see Fig.~s\ref{fig:CNN_analysis_diagram}). The length of these two vectors is the number of comments $N$, usually few thousands. The output is the resulted population parameters $\alpha$ and $\beta$ and their error margin. 

The initial value of $\gamma_{init}$ is taken using the probability for a voter to be in favor of a comment - $P^{com}_{\uparrow}$, which is a measurable parameter given by the ratio of like votes to the total votes:
\begin {equation}\label{Pup player1}
P^{com}_{\uparrow}=\frac{\sum_i\uparrow_i}{\sum_i\uparrow_i+\sum_i\downarrow_i}.
\end {equation}
Taking into account (\ref{eq:comdown}) and $\alpha=\beta=1-\gamma$:
\begin {equation}\label{Pup player2}
P^{com}_\uparrow=\gamma_{init}^2 + (1-\gamma_{init})^2.
\end {equation}
The initial value $\gamma_{init}$ is always chosen to be $>0.5$.

Then one proceeds:

\begin{enumerate}
	\item \textbf{Initialization} 
	\begin{enumerate}
		\item Choose value for the number of voters' threshold: $T$. Start with $T = 10$. 
		\item From now on, consider only comments above the threshold: $\uparrow_i+\downarrow_i > T$.
		\item Define the initial value of the mean vote $\gamma$ by solving the equation: 
		\begin {equation} \nonumber
		\frac{\sum_i{\uparrow_i}}{\sum_i{\uparrow_i}+\sum_i{\downarrow_i}}= {\gamma}^2 + (1-\gamma)^2.
		\end {equation}
		Take only the solution $\gamma>0.5$.
	\end{enumerate}
\item \textbf{Classification of comments} 
\begin{enumerate}
	\item Order the comments according to their like vote fraction: $\frac{\uparrow_i}{\uparrow_i+\downarrow_i}$.
	\item Divide the comments into two groups with ratio of $\gamma$ and $1-\gamma$ according to their like vote fraction, i.e., for group $L$ take the $\gamma$ comments with the highest like vote fraction and for group $D$ take all other comments.  
	\item Calculate the population characteristic parameters $\alpha$ and $\beta$: 
	\begin{eqnarray} \nonumber
	&& 1- \alpha = \sum\limits_{i \in L}{\frac{\uparrow_i}{\uparrow_i+\downarrow_i}},\\\nonumber
	&& \beta = \sum\limits_{i \in D}{\frac{\uparrow_i}{\uparrow_i+\downarrow_i}}.
	\end{eqnarray}
	\item Calculate the new population mean vote $\gamma$ using the values of $\alpha$ and $\beta$: 
	 \begin {eqnarray}\nonumber
	 \gamma=\frac{1-\beta}{1+\alpha-\beta}.
	 \end {eqnarray}
	 \item Repeat stages (a)-(d) until the values of $\alpha$ and $\beta$ converge. 
\end{enumerate}	
\item \textbf{Analyzing} 
\begin{enumerate}
	\item Increase the threshold for the number of voters $T$ by 1, and repeat stages 1-2. 
	\item End when the number of comments above the threshold $N$ is less than 50. 
	\item The resulted $\alpha$ and $\beta$ are the weighted mean over all permitted thresholds:
		\begin{eqnarray} \nonumber
		&& \alpha = \frac{\sum\limits_{T}{\alpha N}}{\sum\limits_{T}{N}},\\\nonumber
		&& \beta = \frac{\sum\limits_{T}{\beta N}}{\sum\limits_{T}{N}}.
		\end{eqnarray}
	\item The resulting $\sigma_\alpha$ and $\sigma_\beta$ are the equivalent standard deviations over all permitted thresholds.
\end{enumerate}

\end{enumerate}

\begin{figure}[!h]
	\resizebox{0.5\textwidth}{!}{\includegraphics{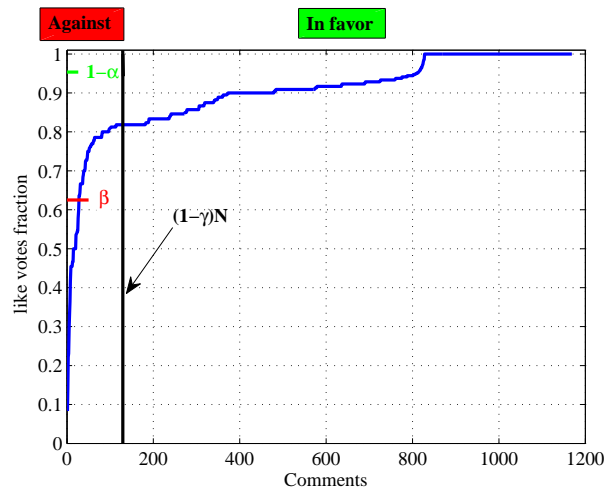}}
	\caption{Like vote fraction distribution for the comments of one CNN article (point 4 in the main article), for threshold value of 10. The black line which is determined by $\gamma$, divides the comments into two groups of in favor and against the subject S. The mean like vote fraction of the against and the in favor groups, equal to $\beta$ and $1-\alpha$ respectively.}
	\label{Smfig1}
\end{figure}

\begin{figure}
	\resizebox{0.5\textwidth}{!}{\includegraphics{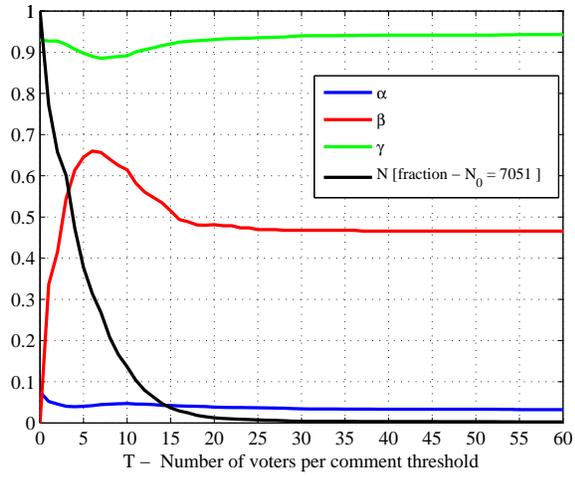}}
	\caption{The sensitivity of the resulting parameters to the threshold value T. The black line represents the number of comments above the threshold and the blue, red, and green lines represent $\alpha,\beta$, and $\gamma$, respectively.}
	\label{Smfig2}
\end{figure}

Fig.~\ref{Smfig1} presents the like vote fraction distribution for the comments of the CNN article announcing the not guilty verdict in the Zimmerman trail (point 4 in the article), for T = 10. The concept of the classification of comments’ procedure and the way the population parameters $\alpha$ and $\beta$ are extracted can be well understood in this presentation. For sensitivity of the model to the value of the threshold $T$ see Fig. \ref{Smfig2}.

\end{document}